\documentclass[12pt]{article}
\usepackage{amsmath,amssymb,amsthm,amsxtra,overpic,bbm,bm,epsfig,ulem}
\usepackage{color}
\usepackage{cite}

\usepackage{hyperref}
\usepackage{url}

\def\thefootnote{\fnsymbol{footnote}}

\begin{document}

\vspace{0.2cm}

\begin{center}
{\large\bf CP violation in light neutrino oscillations and heavy neutrino
decays: \\ a general and explicit seesaw-bridged correlation}
\end{center}

\vspace{0.2cm}

\begin{center}
{\bf Zhi-zhong Xing$^{1,2,3}$}
\footnote{E-mail: xingzz@ihep.ac.cn}
\\
{$^{1}$Institute of High Energy Physics,
Chinese Academy of Sciences, Beijing 100049, China \\
$^{2}$School of Physical Sciences,
University of Chinese Academy of Sciences, Beijing 100049, China \\
$^{3}$Center of High Energy Physics, Peking University, Beijing 100871, China}
\end{center}

\vspace{2cm}
\begin{abstract}
With the help of a block parametrization of the canonical seesaw flavor textures
in terms of the Euler-like rotation angles and CP-violaing phases, we derive a
general and explicit expression for the Jarlskog invariant of CP violation in neutrino
oscillations and compare it with the CP-violating asymmetries of heavy Majorana
neutrino decays within the minimal seesaw framework which contains two right-handed
neutrino fields. Two simplified scenarios are discussed to illustrate how direct
or indirect the correlation between these two types of CP violation can be.
\end{abstract}

\newpage

\def\thefootnote{\arabic{footnote}}
\setcounter{footnote}{0}

\setcounter{equation}{0}
\section{Motivation}

To explain why the three known neutrinos (i.e., the mass eigenstates $\nu^{}_i$
versus the flavor eigenstates $\nu^{}_{\alpha \rm L}$ for $i = 1, 2, 3$ and
$\alpha = e, \mu, \tau$) have nonzero but tiny masses, one may
go beyond the standard model (SM) of particle physics by adding some new degrees
of freedom. In this regard 
the canonical seesaw mechanism~\cite{Minkowski:1977sc,Yanagida:1979as,
GellMann:1980vs,Glashow:1979nm,Mohapatra:1979ia} is not only most economical
but also most natural
%%%%%%%%%%%%%%%%%%%%%%%%%%%%%%%%%%%%%%%%%%%%%%%%%%%%%%%%%%%%%%%%%%%%%%%%%%%%
\footnote{It is also consistent with the essential spirit of effective
field theories for neutrino mass generation at low energies~\cite{Weinberg:1979sa}.},
%%%%%%%%%%%%%%%%%%%%%%%%%%%%%%%%%%%%%%%%%%%%%%%%%%%%%%%%%%%%%%%%%%%%%%%%%%%%
because it just introduces the right-handed neutrino
fields (i.e., the flavor eigenstates $N^{}_{\alpha \rm R}$ versus the mass
eigenstates $N^{}_i$ for $\alpha = e, \mu, \tau$ and $i = 1, 2, 3$) and allows
for the relevant Yukawa interactions and lepton number
violation~\cite{Majorana:1937vz}) which are both harmless to the
fundamentals of the SM. Moreover, the decays of heavy Majorana neutrinos
$N^{}_i$ are CP-violating and thus trigger the leptogenesis
mechanism~\cite{Fukugita:1986hr} to naturally interpret the matter-antimatter
asymmetry observed in the Universe~\cite{ParticleDataGroup:2022pth}.

When confronting the seesaw mechanism with today's experimental and observational
data, however, it is necessary to specify the flavor textures of active and
sterile neutrinos with the help of possible flavor symmetries or
phenomenological assumptions (for the recent reviews with extensive references,
see Refs.~\cite{Xing:2020ijf,Feruglio:2019ybq}). An alternative way, which
is independent of any details of model building, is to fully parametrize the
seesaw flavor textures in terms of the Euler-like rotation angles and CP-violating
phases~\cite{Xing:2007zj,Xing:2011ur}. In particular, one may adopt two
$3\times 3$ unitary submatrices to describe the respective flavor mixing
effects in the presumably isolated active and sterile neutrino sectors, and
utilize a $6\times 6$ non-unitary matrix to characterize the strengths of the
corresponding Yukawa interactions and hence bridge the gap between the two
flavor blocks~\cite{Xing:2011ur}. Such a {\it block parametrization} can not
only reflect the salient dynamical features of
the seesaw mechanism but also make it possible to derive general and explicit
expressions of all the observable quantities of massive neutrinos and thus
reveal their intrinsic correlations in an unprecedentedly transparent manner.

In this work let us show the usefulness of this block parametrization by
deriving a general and explicit expression for the Jarlskog
invariant ${\cal J}^{}_\nu$~\cite{Jarlskog:1985ht,Wu:1985ea,Cheng:1986in}
of leptonic CP violation in neutrino oscillations
and comparing it with the CP-violating asymmetries $\varepsilon^{}_{\alpha i}$
between the lepton-number-violating decays of $N^{}_i$ and their CP-conjugated
processes within the minimal seesaw framework which contains only two right-handed
neutrino fields~\cite{Kleppe:1995zz,Ma:1998zg,Frampton:2002qc}. Our motivation of
doing so is two-fold: (1) the T2K long-baseline neutrino oscillation experiment
has recently ruled out ${\cal J}^{}_\nu = 0$ at the $2\sigma$
level~\cite{T2K:2023smv}, a very encouraging news for the unremitting efforts
to search for CP violation in the lepton sector; (2) the number of free flavor
parameters in the minimal seesaw mechanism is {\it seven} less than that in the
standard seesaw mechanism, making the analytical results of ${\cal J}^{}_\nu$ and
$\varepsilon^{}_{\alpha i}$ less lengthy and thus more instructive. Note that the
minimal seesaw mechanism is still a viable benchmark seesaw framework today and
has attracted a lot of attention in neutrino phenomenology
(for a recent review with extensive references, see Ref.~\cite{Xing:2020ald}).

The remaining parts of this paper are organized as follows. In section 2 we
briefly describe the block parametrization of the active and sterile flavor
textures in the seesaw mechanism. Section 3 is devoted to deriving the
expression of ${\cal J}^{}_\nu$ by taking account of the minimal seesaw
framework, and section 4 compares the dependence of ${\cal J}^{}_\nu$
on the original CP phases with that of $\varepsilon^{}_{\alpha i}$
on the same set of CP phases. In section 5 we discuss two simplified
scenarios of the minimal seesaw mechanism to illustrate how ${\cal J}^{}_\nu$
and $\varepsilon^{}_{\alpha i}$ can be directly or indirectly correlated
with each other in a more transparent way.

\setcounter{equation}{0}
\section{The block parametrization}

Before spontaneous electroweak symmetry breaking, the gauge- and Lorentz-invariant
neutrino mass terms in the canonical seesaw framework can be written
as~\cite{Xing:2023adc}
\begin{eqnarray}
-{\cal L}^{}_{\rm seesaw} \hspace{-0.2cm} & = & \hspace{-0.2cm}
\overline{\ell^{}_{\rm L}} \hspace{0.05cm} Y^{}_\nu \widetilde{H} N^{}_{\rm R}
+ \frac{1}{2} \hspace{0.05cm}
\overline{(N^{}_{\rm R})^c} \hspace{0.05cm} M^{}_{\rm R} N^{}_{\rm R}
+ {\rm h.c.}
\nonumber \\
\hspace{-0.2cm} & = & \hspace{-0.2cm}
\frac{1}{2} \hspace{0.05cm}
\overline{\big[\begin{matrix} \nu^{}_{\rm L} & (N^{}_{\rm R})^c\end{matrix}
\big]} \left(\begin{matrix} {\bf 0} & Y^{}_\nu \phi^{0*} \cr
Y^T_\nu \phi^{0*} & M^{}_{\rm R} \end{matrix}\right)
\left[\begin{matrix} (\nu^{}_{\rm L})^c \cr N^{}_{\rm R} \end{matrix}\right]
- \overline{l^{}_{\rm L}} \hspace{0.05cm} Y^{}_\nu N^{}_{\rm R} \phi^-
+ {\rm h.c.} \; , \hspace{0.5cm}
\label{1}
%     (1)
\end{eqnarray}
in which $\ell^{}_{\rm L} = \big(\begin{matrix} \nu^{}_{\rm L} & l^{}_{\rm L}
\end{matrix}\big)^T$ is the leptonic $\rm SU(2)^{}_{\rm L}$ doublet of
the SM with $\nu^{}_{\rm L} = \big(\begin{matrix} \nu^{}_{e \rm L} &
\nu^{}_{\mu \rm L} & \nu^{}_{\tau \rm L}\end{matrix}\big)^T$ and
$l^{}_{\rm L} = \big(\begin{matrix} l^{}_{e \rm L} & l^{}_{\mu \rm L} &
l^{}_{\tau \rm L}\end{matrix}\big)^T$ standing respectively for the column
vectors of the left-handed neutrino and charged lepton fields,
$\widetilde{H} \equiv {\rm i} \sigma^{}_2 H^*$ with
$H = \big(\begin{matrix} \phi^+ & \phi^0\end{matrix}\big)^T$ denoting the Higgs
doublet of the SM and $\sigma^{}_2$ being the second Pauli matrix,
$N^{}_{\rm R} = \big(\begin{matrix} N^{}_{e \rm R} & N^{}_{\mu \rm R}
& N^{}_{\tau \rm R}\end{matrix}\big)^T$ represents the column vector of the
right-handed neutrino fields which are the $\rm SU(2)^{}_{\rm L}$
singlets, $(\nu^{}_{\rm L})^c \equiv {\cal C} \overline{\nu^{}_{\rm L}}^T$ and
$(N^{}_{\rm R})^c \equiv {\cal C} \overline{N^{}_{\rm R}}^T$
with $\cal C$ being the charge-conjugation matrix, $Y^{}_\nu$ denotes the Yukawa
coupling matrix of neutrinos, and $M^{}_{\rm R}$ is the symmetric right-handed
neutrino mass matrix. Eq.~(\ref{1}) tells us that the left- and right-handed neutrino
fields are bridged by their Yukawa couplings to the neutral component of the
Higgs doublet at the seesaw scale. After the Higgs potential of the SM is minimized
at $\langle H\rangle \equiv \langle 0|H|0\rangle = v/\sqrt{2}$ with a
special direction characterized by $\langle \phi^\pm\rangle = 0$ and
$\langle \phi^0\rangle = v/\sqrt{2}$, where $v \simeq 246 ~{\rm GeV}$ is the vacuum
expectation value of the Higgs field, the $\rm SU(2)^{}_{\rm L} \times U(1)^{}_{\rm Y}$
gauge symmetry is spontaneously broken and consequently all the particles coupled to
$\phi^0$ or $\phi^{0 *}$ acquire their nonzero masses. In this case the terms
associated with $\phi^{\pm}$ in Eq.~(\ref{1}) will disappear, and those associated
with $\phi^{0}$ or $\phi^{0 *}$ will become the so-called Dirac mass matrix
$M^{}_{\rm D} \equiv Y^{}_\nu v/\sqrt{2}$ or its transposed form.

To clearly see how nonzero but tiny masses of the three active neutrinos arise in
this mechanism after electroweak symmetry breaking,
let us diagonalize the symmetric $6\times 6$ matrix in Eq.~(\ref{1})
as follows:
\begin{eqnarray}
\mathbb{U}^\dagger \left ( \begin{matrix} {\bf 0} & M^{}_{\rm D}
\cr M^T_{\rm D} & M^{}_{\rm R} \cr \end{matrix} \right )
\mathbb{U}^* = \left( \begin{matrix} D^{}_\nu & {\bf 0} \cr
{\bf 0} & D^{}_N \cr \end{matrix} \right) \; ,
\label{2}
%     (2)
\end{eqnarray}
where $\mathbb{U}$ is a $6\times 6$ unitary matrix, and the diagonal, real and
positive $3\times 3$ matrices $D^{}_\nu$ and $D^{}_N$ are defined as
$D^{}_\nu \equiv {\rm Diag}\big\{m^{}_1, m^{}_2, m^{}_3\big\}$ and
$D^{}_N \equiv {\rm Diag}\big\{M^{}_1, M^{}_2, M^{}_3 \big\}$ with $m^{}_i$ and
$M^{}_i$ (for $i = 1, 2, 3$) being the respective masses of light and heavy Majorana
neutrinos. We find that decomposing $\mathbb{U}$ into the product of the following
three $6 \times 6$ matrices can help a lot to understand the salient dynamical
features of the canonical seesaw mechanism:
\begin{eqnarray}
\mathbb{U} = \left( \begin{matrix}
I & {\bf 0} \cr {\bf 0} & U^{\prime}_0 \cr \end{matrix} \right)
\left( \begin{matrix} A & R \cr S & B \cr \end{matrix} \right)
\left( \begin{matrix} U^{}_0 & {\bf 0}
\cr {\bf 0} & I \cr \end{matrix} \right) \; ,
\label{3}
%     (3)
\end{eqnarray}
where $I$ denotes the identity matrix, $U^{}_0$ and $U^\prime_0$ are the $3\times 3$
unitary matrices responsible respectively for the primary flavor mixing effects in
the active and sterile neutrino sectors, and $A$, $B$, $R$ and $S$ are the $3\times 3$
non-unitary matrices characterizing the interplay between these two
sectors with the help of the Yukawa interactions~\cite{Xing:2011ur}.
So switching off the Yukawa interactions will immediately lead us to
$D^{}_\nu = {\bf 0}$ or equivalently $m^{}_1 = m^{}_2 = m^{}_3 = 0$, in which
case $U^{}_0$ will lose its physical meaning. Eqs.~(\ref{2}) and (\ref{3}) allow us to
derive out the exact seesaw relationship between the light (active) and heavy
(sterile) Majorana neutrino masses:
\begin{eqnarray}
\left(A U^{}_0\right) D^{}_\nu \left(A U^{}_0\right)^T + R D^{}_N R^T = {\bf 0} \; ,
\label{4}
%     (4)
\end{eqnarray}
in which $A$ and $R$ are also correlated with each other via
$A A^\dagger + R R^\dagger = I$ as a consequence of the unitarity of $\mathbb{U}$.
Note that $U \equiv A U^{}_0$ is just the well-known $3\times 3$
Pontecorvo-Maki-Nakagawa-Sakata (PMNS)
matrix describing flavor mixing and CP violation of the light Majorana neutrinos~\cite{Pontecorvo:1957cp,Maki:1962mu,Pontecorvo:1967fh}, while $R$ is
its analogue measuring the strengths of weak charged-current interactions of the
heavy Majorana neutrinos in the seesaw framework:
\begin{eqnarray}
-{\cal L}^{}_{\rm cc} = \frac{g}{\sqrt{2}} \hspace{0.1cm}
\overline{\big(\begin{matrix} e & \mu & \tau\end{matrix}\big)^{}_{\rm L}}
\hspace{0.1cm} \gamma^\mu \left[ U \left( \begin{matrix} \nu^{}_{1}
\cr \nu^{}_{2} \cr \nu^{}_{3} \cr\end{matrix} \right)^{}_{\hspace{-0.08cm} \rm L}
+ R \left(\begin{matrix} N^{}_{1} \cr N^{}_{2} \cr N^{}_{3}
\cr\end{matrix}\right)^{}_{\hspace{-0.08cm} \rm L} \hspace{0.05cm} \right]
W^-_\mu + {\rm h.c.} \; ,
\label{5}
%     (5)
\end{eqnarray}
where the charged lepton fields are in their mass eigenstates too. Of course,
$U U^\dagger + R R^\dagger = I$ holds, and hence the deviation of $R$ from
$\bf 0$ (or the deviation of $A$ from $I$) characterizes the departure of $U$
from its unitary limit $U^{}_0$. One should keep in mind that both $D^{}_\nu$
and $U^{}_0$ are the consequences of the seesaw mechanism, simply because of
\begin{eqnarray}
U^{}_0 D^{}_\nu U^T_0 =  \left({\rm i} \hspace{0.03cm} A^{-1} R\right) D^{}_N
\left({\rm i} \hspace{0.03cm} A^{-1} R\right)^T \; .
\label{6}
%     (6)
\end{eqnarray}
Namely, the {\it derivational} seesaw parameters on the left-hand side of
Eq.~(\ref{6}) can be determined by the {\it original} seesaw parameters on
the right-handed of Eq.~(\ref{6}). If $M^{}_\nu \equiv U^{}_0 D^{}_\nu U^T_0$
is defined as the effective mass matrix for the three active Majorana neutrinos,
one may easily arrive at the more familiar seesaw formula
$M^{}_\nu \simeq - M^{}_{\rm D} M^{-1}_{\rm R} M^T_{\rm D}$ from Eq.~(\ref{6})
in the approximation of $A \simeq B \simeq I$ (i.e., $M^{}_{\rm D} \simeq
R D^{}_N U^{\prime T}_0$ and $M^{}_{\rm R} \simeq U^\prime_0 D^{}_N U^{\prime T}_0$
hold in this case~\cite{Xing:2023adc}). So Eq.~(\ref{6}) provides a unique
way to constrain the seesaw parameter space by using the available experimental
data on three active neutrinos.

To write out the explicit expressions of $U^{}_0$, $A$ and $R$ in
terms of the Euler-like rotation angles and CP phases, we first introduce
fifteen $6\times 6$ unitary matrices of the form $O^{}_{ij}$ (for
$1 \leq i < j \leq 6$) to realize the block parametrization of
$\mathbb{U}$. The $(i, i)$ and $(j, j)$ elements of $O^{}_{ij}$
are both equal to $c^{}_{ij} \equiv \cos\theta^{}_{ij}$ with $\theta^{}_{ij}$
being a flavor mixing angle and lying in the first quadrant, its other diagonal
entries are all equal to one, its $(i, j)$ and $(j, i)$ elements are given
respectively by $\hat{s}^{*}_{ij} \equiv e^{-{\rm i}\delta^{}_{ij}}
\sin\theta^{}_{ij}$ and $-\hat{s}^{}_{ij} \equiv -e^{{\rm i}\delta^{}_{ij}}
\sin\theta^{}_{ij}$ with $\delta^{}_{ij}$ being a CP-violating phase, and its
other off-diagonal entries are all equal to zero. Then we group these matrices
as follows~\cite{Xing:2011ur}:
\begin{eqnarray}
\left( \begin{matrix} U^{}_0 & 0 \cr 0 & I \cr \end{matrix} \right)
\hspace{-0.2cm} & = & \hspace{-0.2cm} O^{}_{23} O^{}_{13} O^{}_{12} \; ,
\nonumber \\
\left( \begin{matrix} I & 0 \cr 0 & U^{\prime}_0 \cr \end{matrix} \right)
\hspace{-0.2cm} & = & \hspace{-0.2cm} O^{}_{56} O^{}_{46} O^{}_{45} \; ,
\nonumber \\
\left( \begin{matrix} A & R \cr S & B \cr \end{matrix} \right)
\hspace{-0.2cm} & = & \hspace{-0.2cm} O^{}_{36} O^{}_{26} O^{}_{16}
O^{}_{35} O^{}_{25} O^{}_{15} O^{}_{34} O^{}_{24} O^{}_{14} \; , \hspace{0.7cm}
\label{7}
%     (7)
\end{eqnarray}
so as to describe the active flavor sector, the sterile flavor sector and
the interplay between them, respectively, in a way consistent with the salient
features of the seesaw mechanism. The results of $U^{}_0$,
$U^{\prime}_0$, $A$, $R$, $S$ and $B$ have been given in
Refs.~\cite{Xing:2020ijf,Xing:2011ur}. Here we only quote
\begin{eqnarray}
U^{}_0 = \left( \begin{matrix} c^{}_{12} c^{}_{13} & \hat{s}^*_{12}
c^{}_{13} & \hat{s}^*_{13} \cr
-\hat{s}^{}_{12} c^{}_{23} -
c^{}_{12} \hat{s}^{}_{13} \hat{s}^*_{23} & c^{}_{12} c^{}_{23} -
\hat{s}^*_{12} \hat{s}^{}_{13} \hat{s}^*_{23} & c^{}_{13}
\hat{s}^*_{23} \cr
\hat{s}^{}_{12} \hat{s}^{}_{23} - c^{}_{12}
\hat{s}^{}_{13} c^{}_{23} & -c^{}_{12} \hat{s}^{}_{23} -
\hat{s}^*_{12} \hat{s}^{}_{13} c^{}_{23} & c^{}_{13} c^{}_{23}
\cr \end{matrix} \right) \; ,
\label{8}
%     (8)
\end{eqnarray}
and
\begin{eqnarray}
A \hspace{-0.2cm} & = & \hspace{-0.2cm}
\left( \begin{matrix} c^{}_{14} c^{}_{15} c^{}_{16} & 0 & 0
\cr \vspace{-0.45cm} \cr
\begin{array}{l} -c^{}_{14} c^{}_{15} \hat{s}^{}_{16} \hat{s}^*_{26} -
c^{}_{14} \hat{s}^{}_{15} \hat{s}^*_{25} c^{}_{26} \\
-\hat{s}^{}_{14} \hat{s}^*_{24} c^{}_{25} c^{}_{26} \end{array} &
c^{}_{24} c^{}_{25} c^{}_{26} & 0 \cr \vspace{-0.45cm} \cr
\begin{array}{l} -c^{}_{14} c^{}_{15} \hat{s}^{}_{16} c^{}_{26} \hat{s}^*_{36}
+ c^{}_{14} \hat{s}^{}_{15} \hat{s}^*_{25} \hat{s}^{}_{26} \hat{s}^*_{36} \\
- c^{}_{14} \hat{s}^{}_{15} c^{}_{25} \hat{s}^*_{35} c^{}_{36} +
\hat{s}^{}_{14} \hat{s}^*_{24} c^{}_{25} \hat{s}^{}_{26}
\hat{s}^*_{36} \\
+ \hat{s}^{}_{14} \hat{s}^*_{24} \hat{s}^{}_{25} \hat{s}^*_{35}
c^{}_{36} - \hat{s}^{}_{14} c^{}_{24} \hat{s}^*_{34} c^{}_{35}
c^{}_{36} \end{array} &
\begin{array}{l} -c^{}_{24} c^{}_{25} \hat{s}^{}_{26} \hat{s}^*_{36} -
c^{}_{24} \hat{s}^{}_{25} \hat{s}^*_{35} c^{}_{36} \\
-\hat{s}^{}_{24} \hat{s}^*_{34} c^{}_{35} c^{}_{36} \end{array} &
c^{}_{34} c^{}_{35} c^{}_{36} \cr \end{matrix} \right) \; ,
\nonumber \\
R \hspace{-0.2cm} & = & \hspace{-0.2cm}
\left( \begin{matrix} \hat{s}^*_{14} c^{}_{15} c^{}_{16} &
\hat{s}^*_{15} c^{}_{16} & \hat{s}^*_{16} \cr \vspace{-0.45cm} \cr
\begin{array}{l} -\hat{s}^*_{14} c^{}_{15} \hat{s}^{}_{16} \hat{s}^*_{26} -
\hat{s}^*_{14} \hat{s}^{}_{15} \hat{s}^*_{25} c^{}_{26} \\
+ c^{}_{14} \hat{s}^*_{24} c^{}_{25} c^{}_{26} \end{array} & -
\hat{s}^*_{15} \hat{s}^{}_{16} \hat{s}^*_{26} + c^{}_{15}
\hat{s}^*_{25} c^{}_{26} & c^{}_{16} \hat{s}^*_{26} \cr \vspace{-0.45cm} \cr
\begin{array}{l} -\hat{s}^*_{14} c^{}_{15} \hat{s}^{}_{16} c^{}_{26}
\hat{s}^*_{36} + \hat{s}^*_{14} \hat{s}^{}_{15} \hat{s}^*_{25}
\hat{s}^{}_{26} \hat{s}^*_{36} \\ - \hat{s}^*_{14} \hat{s}^{}_{15}
c^{}_{25} \hat{s}^*_{35} c^{}_{36} - c^{}_{14} \hat{s}^*_{24}
c^{}_{25} \hat{s}^{}_{26}
\hat{s}^*_{36} \\
- c^{}_{14} \hat{s}^*_{24} \hat{s}^{}_{25} \hat{s}^*_{35}
c^{}_{36} + c^{}_{14} c^{}_{24} \hat{s}^*_{34} c^{}_{35} c^{}_{36}
\end{array} &
\begin{array}{l} -\hat{s}^*_{15} \hat{s}^{}_{16} c^{}_{26} \hat{s}^*_{36}
- c^{}_{15} \hat{s}^*_{25} \hat{s}^{}_{26} \hat{s}^*_{36} \\
+c^{}_{15} c^{}_{25} \hat{s}^*_{35} c^{}_{36} \end{array} &
c^{}_{16} c^{}_{26} \hat{s}^*_{36} \cr \end{matrix} \right) \; , \hspace{0.5cm}
\label{9}
%     (9)
\end{eqnarray}
which are relevant to the seesaw formula in Eq.~(\ref{4}) or Eq.~(\ref{6}).
It is now clear that the nine flavor parameters associated with the three
light Majorana neutrinos (i.e., $m^{}_i$, $\theta^{}_{ij}$ and $\delta^{}_{ij}$
for $i, j = 1, 2, 3$ and $i < j$) can all be determined by the eighteen
original seesaw parameters (i.e., $M^{}_i$, $\theta^{}_{ij}$ and
$\delta^{}_{ij}$ for $i = 1, 2, 3$ and $j = 4, 5, 6$)
%%%%%%%%%%%%%%%%%%%%%%%%%%%%%%%%%%%%%%%%%%%%%%%%%%%%%%%%%%%%%%%%%%%%%%%%%
\footnote{Note that three of the nine phase parameters (or their independent
combinations) appearing in $A$ and $R$ can always be rotated away through a
proper redefinition of the phases of three charged lepton fields, as one can
easily see from Eq.~(\ref{5}). So we are actually left with six nontrivial
CP-violating phases in the seesaw framework.}
%%%%%%%%%%%%%%%%%%%%%%%%%%%%%%%%%%%%%%%%%%%%%%%%%%%%%%%%%%%%%%%%%%%%%%%%%
through Eq.~(\ref{6}).

A careful analysis of those currently available electroweak precision
measurements and neutrino oscillation data has put some rather strong constraints
on possible non-unitarity of the PMNS matrix $U$ --- the latter is expected to
be of ${\cal O}\left(10^{-3}\right)$ or smaller in magnitude~\cite{Antusch:2006vwa,
Antusch:2009gn,Blennow:2016jkn,Hu:2020oba,Wang:2021rsi,Blennow2023}. In this case
we find that the nine active-sterile flavor mixing angles in $A$ and $R$ must be
smaller than ${\cal O}(10^{-1.5})$. Such a phenomenological
observation means that $A \simeq I$ and $U \simeq U^{}_0$, together with
\begin{eqnarray}
A^{-1} R \hspace{-0.2cm} & = & \hspace{-0.2cm}
\left( \begin{matrix} \hat{t}^{*}_{14} & c^{-1}_{14} \hat{t}^{*}_{15} &
c^{-1}_{14} c^{-1}_{15} \hat{t}^{*}_{16}
\cr \vspace{-0.45cm} \cr
c^{-1}_{14} \hat{t}^{*}_{24} &
\hat{t}^{}_{14} \hat{t}^{*}_{15} \hat{t}^{*}_{24} + c^{-1}_{15} c^{-1}_{24}
\hat{t}^{*}_{25} &
\begin{array}{l}
+ \hat{t}^{}_{14} c^{-1}_{15} \hat{t}^{*}_{16} \hat{t}^{*}_{24}
+ \hat{t}^{}_{15} \hat{t}^{*}_{16} c^{-1}_{24} \hat{t}^{*}_{25} \\
+ c^{-1}_{16} c^{-1}_{24} c^{-1}_{25} \hat{t}^{*}_{26} \end{array}
\cr \vspace{-0.45cm} \cr
c^{-1}_{14} c^{-1}_{24} \hat{t}^{*}_{34} &
\begin{array}{l}
+ \hat{t}^{}_{14} \hat{t}^{*}_{15} c^{-1}_{24} \hat{t}^{*}_{34}
+ c^{-1}_{15} \hat{t}^{}_{24} \hat{t}^{*}_{25} \hat{t}^{*}_{34} \\
+ c^{-1}_{15} c^{-1}_{25} c^{-1}_{34} \hat{t}^{*}_{35}
\end{array} &
\begin{array}{l}
+ \hat{t}^{}_{14} c^{-1}_{15} \hat{t}^{*}_{16} c^{-1}_{24} \hat{t}^{*}_{34}
+ \hat{t}^{}_{15} \hat{t}^{*}_{16} \hat{t}^{}_{24} \hat{t}^{*}_{25}
\hat{t}^{*}_{34} \\
+ \hat{t}^{}_{15} \hat{t}^{*}_{16} c^{-1}_{25} c^{-1}_{34} \hat{t}^{*}_{35}
+ c^{-1}_{16} \hat{t}^{}_{24} c^{-1}_{25} \hat{t}^{*}_{26} \hat{t}^{*}_{34} \\
+ c^{-1}_{16} \hat{t}^{}_{25} \hat{t}^{*}_{26} c^{-1}_{34} \hat{t}^{*}_{35}
+ c^{-1}_{16} c^{-1}_{26} c^{-1}_{34} c^{-1}_{35} \hat{t}^{*}_{36}
\end{array}
\cr \end{matrix} \right) \; \hspace{0.2cm}
\nonumber \\
\hspace{-0.2cm} & \simeq & \hspace{-0.2cm}
\left(\begin{matrix} \hat{s}^*_{14} & \hat{s}^*_{15} & \hat{s}^*_{16} \cr
\hat{s}^*_{24} & \hat{s}^*_{25} & \hat{s}^*_{26} \cr
\hat{s}^*_{34} & \hat{s}^*_{35} & \hat{s}^*_{36} \cr \end{matrix}\right)
+ {\cal O}(s^3_{ij}) + \cdots
\label{10}
%     (10)
\end{eqnarray}
with $\hat{t}^{}_{ij} \equiv e^{{\rm i} \delta^{}_{ij}} \tan\theta^{}_{ij}$
being defined, should be a reliable approximation
(i.e., $c^{}_{ij} \simeq 1$ for sufficiently small
$\theta^{}_{ij}$ with $i = 1, 2, 3$ and $j = 4, 5, 6$).
Hence the flavor texture of $M^{}_\nu$
can be easily reconstructed by using the approximate seesaw relation
$M^{}_\nu \equiv U^{}_0 D^{}_\nu U^T_0 \simeq - R D^{}_N R^T$, from which
the three neutrino masses, three flavor mixing angles and three
CP-violating phases in $D^{}_\nu$ and $U^{}_0$ are consequently extractable.
In particular, one may calculate the effect of leptonic CP violation in
neutrino oscillations by using the original seesaw parameters $M^{}_i$,
$\theta^{}_{ij}$ and $\delta^{}_{ij}$ (for $i = 1, 2, 3$ and $j = 4, 5, 6$).
We are going to explicitly demonstrate this interesting point in the next section.

\section{The Jarlskog invariant ${\cal J}^{}_\nu$}

If the tiny non-unitarity of the $3\times 3$ PMNS flavor mixing
matrix $U$ is neglected within the canonical seesaw framework (i.e., 
$U \simeq U^{}_0$ in the safe approximation of $A \simeq I$),
then the effect of leptonic CP violation in neutrino oscillations
will be uniquely characterized by a universal Jarlskog invariant ${\cal J}^{}_\nu$
which can be defined through~\cite{Jarlskog:1985ht,Wu:1985ea,Cheng:1986in}
\begin{eqnarray}
{\rm Im}\left[\left(U^{}_0\right)^{}_{\alpha i}
\left(U^{}_0\right)^{}_{\beta j} \left(U^{}_0\right)^{*}_{\alpha j}
\left(U^{}_0\right)^{*}_{\beta i} \right] =
{\cal J}^{}_\nu \sum_\gamma \epsilon^{}_{\alpha\beta\gamma} \sum_k
\epsilon^{}_{ijk} \; ,
\label{11}
%     (11)
\end{eqnarray}
in which $\epsilon^{}_{\alpha\beta\gamma}$ and $\epsilon^{}_{ijk}$ denote the
three-dimensional Levi-Civita symbols, and the Greek and Latin subscripts run
respectively over $(e, \mu, \tau)$ and $(1, 2, 3)$. Given the unitarity of
$U^{}_0$ and the definition of $M^{}_\nu \equiv U^{}_0 D^{}_\nu U^T_0$,
a lengthy but straightforward calculation leads us to
\begin{eqnarray}
{\cal J}^{}_\nu = \prod_{i>j} \Delta^{-1}_{ij} \hspace{0.08cm}
{\rm Im}\left[\left(M^{}_\nu M^\dagger_\nu\right)^{}_{e\mu}
\left(M^{}_\nu M^\dagger_\nu\right)^{}_{\mu\tau}
\left(M^{}_\nu M^\dagger_\nu\right)^{}_{\tau e}\right] \; ,
\label{12}
%     (12)
\end{eqnarray}
where $\Delta^{}_{ij} \equiv m^2_i - m^2_j$ (for $i, j = 1, 2, 3$) are
defined
%%%%%%%%%%%%%%%%%%%%%%%%%%%%%%%%%%%%%%%%%%%%%%%%%%%%%%%%%%%%%%%%%%%%%%%%%%%
\footnote{Note that the neutrino mass ordering remains undetermined today (i.e.,
either $m^{}_1 < m^{}_2 < m^{}_3$ or $m^{}_3 < m^{}_2 < m^{}_3$ may be possible,
as indicated by current neutrino oscillation data), but we are absolutely sure
that $\Delta^{}_{31}$ and $\Delta^{}_{32}$ must have the same sign due to the
fact of $\left|\Delta^{}_{31}\right| \simeq \left|\Delta^{}_{32}\right|
\gg \Delta^{}_{21}$~\cite{ParticleDataGroup:2022pth}. So the product
$\Delta^{}_{21} \Delta^{}_{31} \Delta^{}_{32}$ on the right-hand side of
Eq.~(\ref{12}) is actually insensitive to the sign ambiguity
of $\Delta^{}_{31}$ and $\Delta^{}_{32}$.}.
%%%%%%%%%%%%%%%%%%%%%%%%%%%%%%%%%%%%%%%%%%%%%%%%%%%%%%%%%%%%%%%%%%%%%%%%%%%%
A global fit of the present experimental data on solar, atmospheric, reactor
and accelerator neutrino oscillations~\cite{Gonzalez-Garcia:2021dve} provides
us with the best-fit values $\Delta^{}_{21} = 7.42 \times 10^{-5} ~{\rm eV}^2$
and $\Delta^{}_{31} = 2.51 \times 10^{-3} ~{\rm eV}^2$ (normal neutrino mass
ordering) or $\Delta^{}_{32} = -2.49 \times 10^{-3} ~{\rm eV}^2$ (inverted
mass ordering) with relatively small error bars. That is, the product of
$\Delta^{}_{21}$, $\Delta^{}_{31}$ and $\Delta^{}_{32}$ on the right-hand
side of Eq.~(\ref{12}) is already known.

With the help of Eqs.~(\ref{6}) and (\ref{10}), the right-hand side
of Eq.~(\ref{12}) can be expressed in terms of the original seesaw parameters
$M^{}_i$, $\theta^{}_{ij}$ and $\delta^{}_{ij}$ (for $i = 1, 2, 3$ and
$j = 4, 5, 6$). First of all, we obtain
\begin{eqnarray}
\left(M^{}_\nu M^\dagger_\nu\right)^{}_{e\mu} \hspace{-0.2cm} & = & \hspace{-0.2cm}
\hat{s}^*_{14} \hat{s}^{}_{24} A M^2_1 +
\hat{s}^*_{15} \hat{s}^{}_{25} B M^2_2 +
\hat{s}^*_{16} \hat{s}^{}_{26} C M^2_3 +
\left(\hat{s}^*_{14} \hat{s}^{}_{25} X + \hat{s}^*_{15} \hat{s}^{}_{24} X^*\right)
M^{}_1 M^{}_2 \hspace{0.5cm}
\nonumber \\
\hspace{-0.2cm} & & \hspace{-0.2cm}
+ \left(\hat{s}^*_{14} \hat{s}^{}_{26} Y + \hat{s}^*_{16} \hat{s}^{}_{24} Y^*\right)
M^{}_1 M^{}_3 +
\left(\hat{s}^*_{15} \hat{s}^{}_{26} Z + \hat{s}^*_{16} \hat{s}^{}_{25} Z^*\right)
M^{}_2 M^{}_3 \; ,
\nonumber \\
\left(M^{}_\nu M^\dagger_\nu\right)^{}_{\mu\tau} \hspace{-0.2cm} & = & \hspace{-0.2cm}
\hat{s}^*_{24} \hat{s}^{}_{34} A M^2_1 +
\hat{s}^*_{25} \hat{s}^{}_{35} B M^2_2 +
\hat{s}^*_{26} \hat{s}^{}_{36} C M^2_3 +
\left(\hat{s}^*_{24} \hat{s}^{}_{35} X + \hat{s}^*_{25} \hat{s}^{}_{34} X^*\right)
M^{}_1 M^{}_2 \hspace{0.5cm}
\nonumber \\
\hspace{-0.2cm} & & \hspace{-0.2cm}
+ \left(\hat{s}^*_{24} \hat{s}^{}_{36} Y + \hat{s}^*_{26} \hat{s}^{}_{34} Y^*\right)
M^{}_1 M^{}_3 +
\left(\hat{s}^*_{25} \hat{s}^{}_{36} Z + \hat{s}^*_{26} \hat{s}^{}_{35} Z^*\right)
M^{}_2 M^{}_3 \; ,
\nonumber \\
\left(M^{}_\nu M^\dagger_\nu\right)^{}_{\tau e} \hspace{-0.2cm} & = & \hspace{-0.2cm}
\hat{s}^{}_{14} \hat{s}^{*}_{34} A M^2_1 +
\hat{s}^{}_{15} \hat{s}^{*}_{35} B M^2_2 +
\hat{s}^{}_{16} \hat{s}^{*}_{36} C M^2_3 +
\left(\hat{s}^{}_{14} \hat{s}^{*}_{35} X^* + \hat{s}^{}_{15} \hat{s}^{*}_{34} X\right)
M^{}_1 M^{}_2 \hspace{0.5cm}
\nonumber \\
\hspace{-0.2cm} & & \hspace{-0.2cm}
+ \left(\hat{s}^{}_{14} \hat{s}^{*}_{36} Y^* + \hat{s}^{}_{16} \hat{s}^{*}_{34} Y\right)
M^{}_1 M^{}_3 +
\left(\hat{s}^{}_{15} \hat{s}^{*}_{36} Z^* + \hat{s}^{}_{16} \hat{s}^{*}_{35} Z\right)
M^{}_2 M^{}_3 \; ,
\label{13}
%     (13)
\end{eqnarray}
where $A = s^2_{14} + s^2_{24} + s^2_{34}$,
$B = s^2_{15} + s^2_{25} + s^2_{35}$,
$C = s^2_{16} + s^2_{26} + s^2_{36}$, and
\begin{eqnarray}
X \hspace{-0.2cm} & = & \hspace{-0.2cm}
\hat{s}^*_{14} \hat{s}^{}_{15} + \hat{s}^*_{24} \hat{s}^{}_{25} +
\hat{s}^*_{34} \hat{s}^{}_{35} \; ,
\nonumber \\
Y \hspace{-0.2cm} & = & \hspace{-0.2cm}
\hat{s}^*_{14} \hat{s}^{}_{15} + \hat{s}^*_{24} \hat{s}^{}_{26} +
\hat{s}^*_{34} \hat{s}^{}_{36} \; ,
\nonumber \\
Z \hspace{-0.2cm} & = & \hspace{-0.2cm}
\hat{s}^*_{15} \hat{s}^{}_{16} + \hat{s}^*_{25} \hat{s}^{}_{26} +
\hat{s}^*_{35} \hat{s}^{}_{36} \; . \hspace{0.5cm}
\label{14}
%     (14)
\end{eqnarray}
Substituting Eqs.~(\ref{13}) and (\ref{14}) into Eq.~(\ref{12}), one may
calculate the imaginary part of the product of
$\left(M^{}_\nu M^\dagger_\nu\right)^{}_{e\mu}$,
$\left(M^{}_\nu M^\dagger_\nu\right)^{}_{\mu\tau}$ and
$\left(M^{}_\nu M^\dagger_\nu\right)^{}_{\tau e}$. But the analytical
result turns out to be too lengthy to be presented in the present short
note. Instead, we proceed to give a simplified result
within the minimal seesaw mechanism which contains only two right-handed
neutrino fields.

In this case the number of the original seesaw parameters is reduced from
eighteen to eleven. Namely, we are left with two heavy Majorana neutrino
masses ($M^{}_1$ and $M^{}_2$), six active-sterile flavor mixing angles
($\theta^{}_{ij}$ for $i = 1, 2, 3$ and $j = 4, 5$) and three CP-violating
phases (i.e., three of $\delta^{}_{ij}$ for $i = 1, 2, 3$ and $j = 4, 5$,
or their independent combinations) in the minimal seesaw framework. The
corresponding derivational parameters for the light Majorana neutrinos are
seven, including two nonzero masses ($m^{}_2$ and $m^{}_3$ in the normal mass
ordering case, or $m^{}_1$ and $m^{}_2$ in the inverted mass ordering case),
three flavor mixing angles ($\theta^{}_{ij}$ for $ij = 12, 13, 23$) and two
nontrivial CP-violating phases (i.e., two of $\delta^{}_{ij}$ for
$ij = 12, 13, 23$, or their proper combinations). Here let us simply switch
off the parameters $M^{}_3$, $\theta^{}_{i 6}$ and $\delta^{}_{i 6}$ (for
$i = 1, 2, 3$) in Eq.~(\ref{13}) to calculate the Jarlskog invariant
${\cal J}^{}_\nu$ within the minimal seesaw scenario. We arrive at the
analytical result
\begin{eqnarray}
{\rm Im}\left[\left(M^{}_\nu M^\dagger_\nu\right)^{}_{e\mu}
\left(M^{}_\nu M^\dagger_\nu\right)^{}_{\mu\tau}
\left(M^{}_\nu M^\dagger_\nu\right)^{}_{\tau e}\right]
\hspace{-0.2cm} & = & \hspace{-0.2cm}
C^{}_0 \left[ C^{}_{\alpha\beta} \sin \left(\alpha + \beta\right)
+ C^{}_{\beta\gamma} \sin \left(\beta + \gamma\right)
+ C^{}_{\gamma\alpha} \sin \left(\gamma + \alpha\right) \right. \hspace{0.3cm}
\nonumber \\
\hspace{-0.2cm} & & \hspace{-0.2cm}
+ \hspace{0.07cm} C^{\prime}_{\alpha\beta} \sin \left(\alpha - \beta\right)
+ C^{\prime}_{\beta\gamma} \sin \left(\beta - \gamma\right)
+ C^{\prime}_{\gamma\alpha} \sin \left(\gamma - \alpha\right)
\nonumber \\
\hspace{-0.2cm} & & \hspace{-0.2cm}
+ \left. \hspace{-0.05cm} C^{}_{2\alpha} \sin 2\alpha + C^{}_{2\beta}
\sin 2\beta + C^{}_{2\gamma} \sin 2\gamma \right] \; ,
\label{15}
%     (15)
\end{eqnarray}
where $\alpha \equiv \delta^{}_{14} - \delta^{}_{15}$, $\beta \equiv \delta^{}_{24}
- \delta^{}_{25}$ and $\gamma \equiv \delta^{}_{34} - \delta^{}_{35}$, together with
\begin{eqnarray}
C^{}_0 \hspace{-0.2cm} & = & \hspace{-0.2cm}
M^2_1 M^2_2 \left[ s^2_{14} \left( s^2_{25} + s^2_{35}\right)
+ s^2_{24} \left(s^2_{15} + s^2_{35}\right) + s^2_{34} \left(s^2_{15} +
s^2_{25}\right) \right.
\nonumber \\
\hspace{-0.2cm} && \hspace{-0.2cm}
- \left. \hspace{-0.05cm}
2 s^{}_{14} s^{}_{15} s^{}_{24} s^{}_{25} \cos \left(\alpha - \beta\right)
- 2 s^{}_{24} s^{}_{25} s^{}_{34} s^{}_{35} \cos \left(\beta - \gamma\right)
- 2 s^{}_{14} s^{}_{15} s^{}_{34} s^{}_{35} \cos \left(\gamma - \alpha\right)
\right] \; , \hspace{0.5cm}
\label{16}
%     (16)
\end{eqnarray}
and
\begin{eqnarray}
C^{}_{2\alpha} \hspace{-0.2cm} & = & \hspace{-0.2cm}
M^{}_1 M^{}_2 s^2_{14} s^2_{15} \left(s^2_{34} s^2_{25} - s^2_{24} s^2_{35}\right)
\; ,
\nonumber \\
C^{}_{2\beta} \hspace{-0.2cm} & = & \hspace{-0.2cm}
M^{}_1 M^{}_2 s^2_{24} s^2_{25} \left(s^2_{14} s^2_{35} - s^2_{34} s^2_{15}\right)
\; ,
\nonumber \\
C^{}_{2\gamma} \hspace{-0.2cm} & = & \hspace{-0.2cm}
M^{}_1 M^{}_2 s^2_{34} s^2_{35} \left(s^2_{24} s^2_{15} - s^2_{14} s^2_{25}\right)
\; ,
\nonumber \\
C^{}_{\alpha\beta} \hspace{-0.2cm} & = & \hspace{-0.2cm}
M^{}_1 M^{}_2 s^{}_{14} s^{}_{15} s^{}_{24} s^{}_{25}
\left(s^2_{14} s^2_{35} - s^2_{34} s^2_{15} + s^2_{34} s^2_{25} - s^2_{24} s^2_{35}\right)
\; ,
\nonumber \\
C^{}_{\beta\gamma} \hspace{-0.2cm} & = & \hspace{-0.2cm}
M^{}_1 M^{}_2 s^{}_{24} s^{}_{25} s^{}_{34} s^{}_{35}
\left(s^2_{14} s^2_{35} - s^2_{34} s^2_{15} + s^2_{24} s^2_{15} - s^2_{14} s^2_{25}\right)
\; ,
\nonumber \\
C^{}_{\gamma\alpha} \hspace{-0.2cm} & = & \hspace{-0.2cm}
M^{}_1 M^{}_2 s^{}_{14} s^{}_{15} s^{}_{34} s^{}_{35}
\left(s^2_{24} s^2_{15} - s^2_{14} s^2_{25} + s^2_{34} s^2_{25} - s^2_{24} s^2_{35}\right)
\; ,
\nonumber \\
C^{\prime}_{\alpha\beta} \hspace{-0.2cm} & = & \hspace{-0.2cm}
\left[ M^2_1 \left(s^2_{14} + s^2_{24} + s^2_{34}\right) s^2_{34}
- M^2_2 \left(s^2_{15} + s^2_{25} + s^2_{35}\right) s^2_{35} \right]
s^{}_{14} s^{}_{15} s^{}_{24} s^{}_{25} \; ,
\nonumber \\
C^{\prime}_{\beta\gamma} \hspace{-0.2cm} & = & \hspace{-0.2cm}
\left[ M^2_1 \left(s^2_{14} + s^2_{24} + s^2_{34}\right) s^2_{14}
- M^2_2 \left(s^2_{15} + s^2_{25} + s^2_{35}\right) s^2_{15} \right]
s^{}_{24} s^{}_{25} s^{}_{34} s^{}_{35} \; ,
\nonumber \\
C^{\prime}_{\gamma\alpha} \hspace{-0.2cm} & = & \hspace{-0.2cm}
\left[ M^2_1 \left(s^2_{14} + s^2_{24} + s^2_{34}\right) s^2_{24}
- M^2_2 \left(s^2_{15} + s^2_{25} + s^2_{35}\right) s^2_{25} \right]
s^{}_{14} s^{}_{15} s^{}_{34} s^{}_{35} \; . \hspace{0.5cm}
\label{17}
%     (17)
\end{eqnarray}
It becomes obvious how the Jarlskog invariant ${\cal J}^{}_\nu$ depends on the three
original CP-violating phases $\alpha$, $\beta$ and $\gamma$. Since these three phase
parameters actually determine the CP-violating effects in those lepton-number-violating
decays of $N^{}_1$ and $N^{}_2$, one may therefore establish a general and explicit
correlation between CP violation in light neutrino oscillations and that in
heavy Majorana neutrino decays within the minimal seesaw framework under discussion.

\section{The CP asymmetries $\varepsilon^{}_{i \alpha}$}

Given that the canonical seesaw scale is expected to be far above the scale of
electroweak symmetry breaking, all the SM particles should be exactly massless
when the heavy Majorana neutrinos $N^{}_i$ decay. In this case one only needs to
calculate the lepton-number-violating decays of $N^{}_i$ into the leptonic doublet
and the Higgs doublet at the one-loop level, so as to determine the CP-violating
asymmetries $\varepsilon^{}_{i \alpha}$ between $N^{}_i \to \ell^{}_\alpha + H$
and their CP-conjugated processes~\cite{Fukugita:1986hr,Luty:1992un,Covi:1996wh,
Plumacher:1996kc,Pilaftsis:1997jf}.
%%%%%%%%%%%%%%%%%%%%%%%%%%%%%%%%%%%%%%%%%%%%%%%%%%%%%%%%%%%%%%%%%%%%%%%%%%%%%%%%%
%\footnote{Note that a tiny mismatch between the mass eigenstates of heavy
%Majorana neutrinos associated with their decays and with the canonical seesaw
%mechanism itself~\cite{Xing:2023adc,Drewes:2013gca,Canetti:2012kh,Drewes:2019byd,
%Chrzaszcz:2019inj}, which only constitutes an extremely small sub-leading
%contribution to the size of $\varepsilon^{}_{i\alpha}$, has been safely neglected
%here.}.
%%%%%%%%%%%%%%%%%%%%%%%%%%%%%%%%%%%%%%%%%%%%%%%%%%%%%%%%%%%%%%%%%%%%%%%%%%%%%%%%%
Namely,
\begin{eqnarray}
\varepsilon^{}_{i \alpha} \hspace{-0.2cm} & \equiv & \hspace{-0.2cm}
\frac{\Gamma({N}^{}_i \to \ell^{}_\alpha + H)
- \Gamma({N}^{}_i \to \overline{\ell^{}_\alpha} +
\overline{H})}{\displaystyle \sum_\alpha \left[\Gamma({N}^{}_i \to
\ell^{}_\alpha + H) + \Gamma({N}^{}_i \to \overline{\ell^{}_\alpha}
+ \overline{H})\right]}
\nonumber \\
\hspace{-0.2cm} & = & \hspace{-0.2cm}
\frac{1}{8\pi \big({\cal Y}^\dagger_\nu {\cal Y}^{}_\nu\big)^{}_{ii}} \sum_{j \neq i}
\left\{ {\rm Im} \left[\big({\cal Y}^*_\nu\big)^{}_{\alpha i}
\big({\cal Y}^{}_\nu\big)^{}_{\alpha j}
\big({\cal Y}^\dagger_\nu {\cal Y}^{}_\nu\big)^{}_{ij}
\hspace{0.05cm} \xi(x^{}_{ji}) + \big({\cal Y}^*_\nu\big)^{}_{\alpha i}
\big({\cal Y}^{}_\nu\big)^{}_{\alpha j}
\big({\cal Y}^\dagger_\nu {\cal Y}^{}_\nu\big)^*_{ij}
\hspace{0.05cm} \zeta(x^{}_{ji}) \right]\right\} \; , \hspace{0.5cm}
\label{18}
%     (18)
\end{eqnarray}
where ${\cal Y}^{}_\nu \equiv Y^{}_\nu U^{\prime *}_0$ is
defined~\cite{Xing:2023adc}, the Latin and Greek subscripts run respectively
over $(1, 2, 3)$ and $(e, \mu, \tau)$,
$\xi(x^{}_{ji}) = \sqrt{x^{}_{ji}} \left\{1 + 1/\left(1 - x^{}_{ji}\right)
+ \left(1 + x^{}_{ji}\right) \ln \left[x^{}_{ji} / \left(1 + x^{}_{ji}\right)
\right] \right\}$ and $\zeta(x^{}_{ji}) = 1/\left(1 - x^{}_{ji}\right)$
with $x^{}_{ji} \equiv {M}^2_j/{M}^2_i$~\cite{Xing:2011zza}.
%%%%%%%%%%%%%%%%%%%%%%%%%%%%%%%%%%%%%%%%%%%%%%%%%%%%%%%%%%%%%%%%%%%%%%%%%%%%
%\footnote{If $x^{}_{ji} \gg 1$ holds, one will arrive at
%$\xi(x^{}_{ji}) \simeq -3 M^{}_i/\left(2 M^{}_i\right)$ and
%$\zeta(x^{}_{ji}) \simeq -M^{2}_i/M^{2}_j$ as a very good approximation.}.
%%%%%%%%%%%%%%%%%%%%%%%%%%%%%%%%%%%%%%%%%%%%%%%%%%%%%%%%%%%%%%%%%%%%%%%%%%%%
The key point of thermal leptogenesis is that a net lepton-antilepton asymmetry
can arise from nonzero $\varepsilon^{}_{i \alpha}$ in the early Universe, and
later on it can be partly converted into a net baryon-antibaryon asymmetry via
the sphaleron interactions~\cite{Fukugita:1986hr,Kuzmin:1985mm,Harvey:1990qw},
providing a qualitatively elegant explanation of why the primordial antibaryons
have essentially disappeared in the Universe.

To see how the CP-violating asymmetries $\varepsilon^{}_{i \alpha}$ are
explicitly dependent upon the original CP phases in the minimal
seesaw mechanism with only two right-handed neutrino fields, let us first
substitute $Y^{}_\nu \simeq R D^{}_N U^{\prime T}_0/\langle \phi^0\rangle$
into ${\cal Y}^{}_\nu \equiv Y^{}_\nu U^{\prime *}_0$ and then substitute the
expression of ${\cal Y}^{}_\nu$ into Eq.~(\ref{18}) in the approximation of
$A \simeq B \simeq I$. After switching off the flavor mixing angles
$\theta^{}_{i 6}$ (for $i = 1, 2, 3$) associated with $N^{}_3$ in
Eqs.~(\ref{9}) and (\ref{10}), we are left with
\begin{eqnarray}
\varepsilon^{}_{1 e} \hspace{-0.2cm} & = & \hspace{-0.2cm}
\frac{M^2_2 s^{}_{14} s^{}_{15}}{8\pi \langle \phi^0\rangle^2
\left(s^2_{14} + s^2_{24} + s^2_{34}\right)}
\Big[\xi(x^{}_{21}) \big[ s^{}_{14} s^{}_{15}
\sin 2\alpha + s^{}_{24} s^{}_{25} \sin \left(\alpha + \beta\right)
+ s^{}_{34} s^{}_{35} \sin\left( \alpha + \gamma\right)\big]
\nonumber \\
\hspace{-0.2cm} & & \hspace{-0.2cm}
+ \hspace{0.05cm} \zeta(x^{}_{21}) \big[ s^{}_{24} s^{}_{25}
\sin\left(\alpha - \beta\right) + s^{}_{34} s^{}_{35}
\sin\left(\alpha - \gamma\right)\big] \Big] \; ,
\nonumber \\
\varepsilon^{}_{1 \mu} \hspace{-0.2cm} & = & \hspace{-0.2cm}
\frac{M^2_2 s^{}_{24} s^{}_{25}}{8\pi \langle \phi^0\rangle^2
\left(s^2_{14} + s^2_{24} + s^2_{34}\right)}
\Big[\xi(x^{}_{21}) \big[ s^{}_{14} s^{}_{15}
\sin\left(\alpha + \beta\right) + s^{}_{24} s^{}_{25} \sin 2\beta
+ s^{}_{34} s^{}_{35} \sin\left( \beta + \gamma\right)\big]
\nonumber \\
\hspace{-0.2cm} & & \hspace{-0.2cm}
+ \hspace{0.05cm} \zeta(x^{}_{21}) \big[ s^{}_{14} s^{}_{15}
\sin\left(\beta - \alpha\right) + s^{}_{34} s^{}_{35}
\sin\left(\beta - \gamma\right)\big] \Big] \; ,
\nonumber \\
\varepsilon^{}_{1 \tau} \hspace{-0.2cm} & = & \hspace{-0.2cm}
\frac{M^2_2 s^{}_{34} s^{}_{35}}{8\pi \langle \phi^0\rangle^2
\left(s^2_{14} + s^2_{24} + s^2_{34}\right)}
\Big[\xi(x^{}_{21}) \big[ s^{}_{14} s^{}_{15}
\sin \left(\alpha + \gamma\right) + s^{}_{24} s^{}_{25}
\sin \left(\beta + \gamma\right) + s^{}_{34} s^{}_{35} \sin 2\gamma \big]
\nonumber \\
\hspace{-0.2cm} & & \hspace{-0.2cm}
+ \hspace{0.05cm} \zeta(x^{}_{21}) \big[ s^{}_{14} s^{}_{15}
\sin\left(\gamma - \alpha\right) + s^{}_{24} s^{}_{25}
\sin\left(\gamma - \beta\right)\big] \Big] \; ,
\label{19}
%     (19)
\end{eqnarray}
for the decays of $N^{}_1$; and
\begin{eqnarray}
\varepsilon^{}_{2 e} \hspace{-0.2cm} & = & \hspace{-0.2cm}
- \frac{M^2_1 s^{}_{14} s^{}_{15}}{8\pi \langle \phi^0\rangle^2
\left(s^2_{15} + s^2_{25} + s^2_{35}\right)}
\Big[ \xi(x^{}_{12}) \big[ s^{}_{14} s^{}_{15}
\sin 2\alpha + s^{}_{24} s^{}_{25} \sin \left(\alpha + \beta\right)
+ s^{}_{34} s^{}_{35} \sin\left( \alpha + \gamma\right)\big]
\nonumber \\
\hspace{-0.2cm} & & \hspace{-0.2cm}
+ \hspace{0.05cm} \zeta(x^{}_{12}) \big[ s^{}_{24} s^{}_{25}
\sin\left(\alpha - \beta\right) + s^{}_{34} s^{}_{35}
\sin\left(\alpha - \gamma\right)\big] \Big] \; ,
\nonumber \\
\varepsilon^{}_{2 \mu} \hspace{-0.2cm} & = & \hspace{-0.2cm}
- \frac{M^2_1 s^{}_{24} s^{}_{25}}{8\pi \langle \phi^0\rangle^2
\left(s^2_{15} + s^2_{25} + s^2_{35}\right)}
\Big[ \xi(x^{}_{12}) \big[ s^{}_{14} s^{}_{15}
\sin\left(\alpha + \beta\right) + s^{}_{24} s^{}_{25} \sin 2\beta
+ s^{}_{34} s^{}_{35} \sin\left( \beta + \gamma\right)\big]
\nonumber \\
\hspace{-0.2cm} & & \hspace{-0.2cm}
+ \hspace{0.05cm} \zeta(x^{}_{12}) \big[ s^{}_{14} s^{}_{15}
\sin\left(\beta - \alpha\right) + s^{}_{34} s^{}_{35}
\sin\left(\beta - \gamma\right)\big] \Big] \; ,
\nonumber \\
\varepsilon^{}_{2 \tau} \hspace{-0.2cm} & = & \hspace{-0.2cm}
- \frac{M^2_1 s^{}_{34} s^{}_{35}}{8\pi \langle \phi^0\rangle^2
\left(s^2_{15} + s^2_{25} + s^2_{35}\right)}
\Big[ \xi(x^{}_{12}) \big[ s^{}_{14} s^{}_{15}
\sin \left(\alpha + \gamma\right) + s^{}_{24} s^{}_{25}
\sin \left(\beta + \gamma\right) + s^{}_{34} s^{}_{35} \sin 2\gamma \big]
\nonumber \\
\hspace{-0.2cm} & & \hspace{-0.2cm}
+ \hspace{0.05cm} \zeta(x^{}_{12}) \big[ s^{}_{14} s^{}_{15}
\sin\left(\gamma - \alpha\right) + s^{}_{24} s^{}_{25}
\sin\left(\gamma - \beta\right)\big] \Big] \; ,
\label{20}
%     (20)
\end{eqnarray}
for the decays of $N^{}_2$, where the phase parameters
$\alpha \equiv \delta^{}_{14} - \delta^{}_{15}$,
$\beta \equiv \delta^{}_{24} - \delta^{}_{25}$ and
$\gamma \equiv \delta^{}_{34} - \delta^{}_{35}$ have been defined as in
Eqs.~(\ref{15}) and (\ref{16}). A sum of $\varepsilon^{}_{i \alpha}$ over
$\alpha$ (for $\alpha = e, \mu, \tau$) immediately leads us to the
flavor-independent CP-violating asymmetries
\begin{eqnarray}
\varepsilon^{}_{1} \hspace{-0.2cm} & = & \hspace{-0.2cm}
+ \frac{M^2_2 \xi(x^{}_{21})}{8\pi \langle \phi^0\rangle^2
\left(s^2_{14} + s^2_{24} + s^2_{34}\right)}
\big[ s^{2}_{14} s^{2}_{15}
\sin 2\alpha + s^2_{24} s^2_{25} \sin 2\beta + s^2_{34} s^2_{35} \sin 2\gamma
\nonumber \\
\hspace{-0.2cm} & & \hspace{-0.2cm}
+ \hspace{0.07cm}
2 s^{}_{14} s^{}_{15} s^{}_{24} s^{}_{25} \sin \left(\alpha + \beta\right)
+ 2 s^{}_{14} s^{}_{15} s^{}_{34} s^{}_{35} \sin \left(\alpha + \gamma\right)
+ 2 s^{}_{24} s^{}_{25} s^{}_{34} s^{}_{35} \sin \left(\beta + \gamma\right)
\big] \; ,
\nonumber \\
\varepsilon^{}_{2} \hspace{-0.2cm} & = & \hspace{-0.2cm}
- \frac{M^2_1 \xi(x^{}_{12})}{8\pi \langle \phi^0\rangle^2
\left(s^2_{15} + s^2_{25} + s^2_{35}\right)}
\big[ s^{2}_{14} s^{2}_{15}
\sin 2\alpha + s^2_{24} s^2_{25} \sin 2\beta + s^2_{34} s^2_{35} \sin 2\gamma
\nonumber \\
\hspace{-0.2cm} & & \hspace{-0.2cm}
+ \hspace{0.07cm}
2 s^{}_{14} s^{}_{15} s^{}_{24} s^{}_{25} \sin \left(\alpha + \beta\right)
+ 2 s^{}_{14} s^{}_{15} s^{}_{34} s^{}_{35} \sin \left(\alpha + \gamma\right)
+ 2 s^{}_{24} s^{}_{25} s^{}_{34} s^{}_{35} \sin \left(\beta + \gamma\right)
\big] \; , \hspace{0.4cm}
\label{21}
%     (21)
\end{eqnarray}
in which the terms associated with the loop functions $\zeta(x^{}_{21})$ and
$\zeta(x^{}_{12})$ in Eqs.~(\ref{19}) and (\ref{20}) have been exactly
cancelled out. It is obvious that the three independent CP phases $\alpha$,
$\beta$ and $\gamma$ appear both in the formula of ${\cal J}^{}_\nu$ and in
the expressions of $\varepsilon^{}_{i \alpha}$ and $\varepsilon^{}_i$ (for
$i = 1, 2$ and $\alpha = e, \mu, \tau$), implying a kind of correlation
between CP violation in light neutrino oscillations (with lepton number
conservation) and that in heavy neutrino decays (with lepton number violation).
In general, however, there
is no proportional relation between these two types of CP-violating effects.
This observation means that $\varepsilon^{}_{i \alpha} = 0$ (or
$\varepsilon^{}_i = 0$) does not necessarily result in ${\cal J}^{}_\nu = 0$,
or vice versa.
%%%%%%%%%%%%%%%%%%%%%%%%%%%%%%%  Table 1  %%%%%%%%%%%%%%%%%%%%%%%%%%%%%%%%%%%%%
\begin{table}[t]
\caption{A brief summary of the dependence of ${\cal J}^{}_\nu$,
$\varepsilon^{}_{i \alpha}$ and $\varepsilon^{}_i$ (for $i = 1, 2$ and
$\alpha = e, \mu, \tau$) on the three CP phases $\alpha$, $\beta$ and
$\gamma$ in the minimal seesaw mechanism, where the symbol ``$\surd$" means
the existence of the $\sin 2\alpha$ term and so on.
\label{Table1}}
\vspace{0cm}
\small
\begin{center}
\begin{tabular}{c|ccc|ccc|ccc} \hline\hline
     & $\sin 2\alpha$ & $\sin 2\beta$ & $\sin 2\gamma$
     & $\sin\left(\alpha + \beta\right)$
     & $\sin\left(\alpha + \gamma\right)$ & $\sin\left(\beta + \gamma\right)$
     & $\sin\left(\alpha - \beta\right)$  & $\sin\left(\beta - \gamma\right)$
     & $\sin\left(\gamma - \alpha\right)$ \\ \hline\hline
%%%%%%%%%%%%%%%%%%%%%%%%%%%%%%%%%%%%%%%%%%%%%%%%%%%%%%%%%%%%%%%%%%%%%%%%
${\cal J}^{}_\nu$ & $\surd$ & $\surd$ & $\surd$ & $\surd$ & $\surd$ & $\surd$
& $\surd$ & $\surd$ & $\surd$ \\ \hline\hline
%%%%%%%%%%%%%%%%%%%%%%%%%%%%%%%%%%%%%%%%%%%%%%%%%%%%%%%%%%%%%%%%%%%%%%%%%%%%%%%
%%%%%%%%%%%%%%%%%%%%%%%%%%%%%%%%%%%%%%%%%%%%%%%%%%%%%%%%%%%%%%%%%%%%%%%%%%%%%%%
$\varepsilon^{}_{1 e}$ & $\surd$ &  &  & $\surd$ & $\surd$ &
& $\surd$ &  & $\surd$ \\ \hline
%%%%%%%%%%%%%%%%%%%%%%%%%%%%%%%%%%%%%%%%%%%%%%%%%%%%%%%%%%%%%%%%%%%%%%%%%%%%%%%
$\varepsilon^{}_{1 \mu}$ &  & $\surd$ &  & $\surd$ &  & $\surd$
& $\surd$ & $\surd$  & \\ \hline
%%%%%%%%%%%%%%%%%%%%%%%%%%%%%%%%%%%%%%%%%%%%%%%%%%%%%%%%%%%%%%%%%%%%%%%%%%%%%%%
$\varepsilon^{}_{1 \tau}$ &  &  & $\surd$ &  & $\surd$ & $\surd$
&  & $\surd$ & $\surd$ \\ \hline
%%%%%%%%%%%%%%%%%%%%%%%%%%%%%%%%%%%%%%%%%%%%%%%%%%%%%%%%%%%%%%%%%%%%%%%%
$\varepsilon^{}_1$ & $\surd$ & $\surd$ & $\surd$ & $\surd$ & $\surd$ & $\surd$
&  &  &  \\ \hline\hline
%%%%%%%%%%%%%%%%%%%%%%%%%%%%%%%%%%%%%%%%%%%%%%%%%%%%%%%%%%%%%%%%%%%%%%%%%%%%%%%
%%%%%%%%%%%%%%%%%%%%%%%%%%%%%%%%%%%%%%%%%%%%%%%%%%%%%%%%%%%%%%%%%%%%%%%%%%%%%%%
$\varepsilon^{}_{2 e}$ & $\surd$ &  &  & $\surd$ & $\surd$ &
& $\surd$ &  & $\surd$ \\ \hline
%%%%%%%%%%%%%%%%%%%%%%%%%%%%%%%%%%%%%%%%%%%%%%%%%%%%%%%%%%%%%%%%%%%%%%%%%%%%%%%
$\varepsilon^{}_{2 \mu}$ &  & $\surd$ &  & $\surd$ &  & $\surd$
& $\surd$ & $\surd$  & \\ \hline
%%%%%%%%%%%%%%%%%%%%%%%%%%%%%%%%%%%%%%%%%%%%%%%%%%%%%%%%%%%%%%%%%%%%%%%%%%%%%%%
$\varepsilon^{}_{2 \tau}$ &  &  & $\surd$ &  & $\surd$ & $\surd$
&  & $\surd$ & $\surd$ \\ \hline
%%%%%%%%%%%%%%%%%%%%%%%%%%%%%%%%%%%%%%%%%%%%%%%%%%%%%%%%%%%%%%%%%%%%%%%%%%%%%%%
$\varepsilon^{}_2$ & $\surd$ & $\surd$ & $\surd$ & $\surd$ & $\surd$ & $\surd$
&  &  &  \\ \hline\hline
%%%%%%%%%%%%%%%%%%%%%%%%%%%%%%%%%%%%%%%%%%%%%%%%%%%%%%%%%%%%%%%%%%%%%%%%%%%%%%%
%%%%%%%%%%%%%%%%%%%%%%%%%%%%%%%%%%%%%%%%%%%%%%%%%%%%%%%%%%%%%%%%%%%%%%%%%%%%%%%
\end{tabular}
\end{center}
\end{table}
%%%%%%%%%%%%%%%%%%%%%%%%%%%%%%%%%%%%%%%%%%%%%%%%%%%%%%%%%%%%%%%%%%%%%%%%%%%%%

Table~\ref{Table1} is a brief summary of the dependence of ${\cal J}^{}_\nu$,
$\varepsilon^{}_{i \alpha}$ and $\varepsilon^{}_i$ (for $i = 1, 2$ and
$\alpha = e, \mu, \tau$) on the three original CP phases $\alpha$, $\beta$ and
$\gamma$ in the minimal seesaw mechanism. Note that the remarkable result of
${\cal J}^{}_\nu$ obtained in Eqs.~(\ref{12}) and (\ref{15})---(\ref{17}) has
been expressed in terms of the original seesaw parameters, and hence the resulting
correlation between ${\cal J}^{}_\nu$ and $\varepsilon^{}_{i \alpha}$ (or
$\varepsilon^{}_i$) is more straightforward and more transparent than that
derived from the basis-independent flavor invariants in the seesaw-based
effective field theory~\cite{Wang:2021wdq,Yu:2021cco,Yu:2022nxj,Yu:2022ttm}.
Our analytical results are expected to be useful for a systematic analysis of
the parameter space of CP violation in the minimal seesaw framework, once more
experimental and observational data are available in the foreseeable future.

\section{Two simplified scenarios}

We consider two special scenarios of the active-sterile flavor mixing
pattern to simplify the expressions of ${\cal J}^{}_\nu$,
$\varepsilon^{}_{i \alpha}$ and $\varepsilon^{}_i$ (for $i = 1, 2$ and
$\alpha = e, \mu, \tau$) obtained above, so as to
arrive at a more direct correlation between the effects of CP violation in
light neutrino oscillations and heavy neutrino decays.

\subsection{\bf Scenario A: two texture zeros}

Taking $\theta^{}_{15} = \theta^{}_{34} = 0$ leads us to a simple but
viable texture of $M^{}_{\rm D} \propto R$ with the vanishing $(1, 2)$ and
$(3, 1)$ entries, as first proposed in Ref.~\cite{Frampton:2002qc}. In
this case only the coefficients $C^{}_0$ and $C^{}_{2\beta}$ in
Eq.~(\ref{15}) are nonzero, and hence we are left with
\begin{eqnarray}
{\cal J}^{}_\nu = \frac{M^3_1 M^3_2}{\Delta^{}_{21} \Delta^{}_{31}
\Delta^{}_{32}} s^2_{14} s^2_{24} s^2_{25} s^2_{35} \left[s^2_{14} s^2_{25}
+ s^2_{14} s^2_{35} + s^2_{24} s^2_{35} \right] \sin 2\beta \; ,
\label{22}
%     (22)
\end{eqnarray}
which depends only on a single phase parameter $\beta$. On the other hand,
we obtain
\begin{eqnarray}
\varepsilon^{}_{1} \hspace{-0.2cm} & = & \hspace{-0.2cm}
\varepsilon^{}_{1 \mu} =
+ \frac{M^2_2 \xi(x^{}_{21}) s^2_{24} s^2_{25}}{8\pi \langle \phi^0\rangle^2
\left(s^2_{14} + s^2_{24} \right)}
\sin 2\beta \; ,
\nonumber \\
\varepsilon^{}_{2} \hspace{-0.2cm} & = & \hspace{-0.2cm}
\varepsilon^{}_{2 \mu} =
- \frac{M^2_1 \xi(x^{}_{12}) s^2_{24} s^2_{25}}{8\pi \langle \phi^0\rangle^2
\left(s^2_{25} + s^2_{35}\right)}
\sin 2\beta \; , \hspace{0.4cm}
\label{23}
%     (23)
\end{eqnarray}
which depend on the same phase parameter $\beta$,
together with $\varepsilon^{}_{1 e} = \varepsilon^{}_{2 e} =
\varepsilon^{}_{1 \tau} = \varepsilon^{}_{2 \tau} = 0$. This scenario can
therefore establish a direct link between CP violation in neutrino
oscillations and that in heavy Majorana neutrino decays (i.e.,
${\cal J}^{}_\nu \propto \varepsilon^{}_1 \propto \varepsilon^{}_2$).
The latter is responsible for the validity of thermal
leptogenesis~\cite{Frampton:2002qc,Xing:2020ald}, so as to interpret
the cosmic matter-antimatter asymmetry of the Universe.

One may wonder whether the direct proportionality between
$J^{}_\nu$ and $\varepsilon^{}_{1,2}$ obtained above will be lost if
the active-sterile flavor mixing angles are not small and hence
the approximation made in Eq.~(\ref{10}) is not reliable anymore.
Within the minimal seesaw framework, the exact expression of $A^{-1} R$
in the assumption of $\theta^{}_{15} = \theta^{}_{34} = 0$ becomes
\begin{eqnarray}
A^{-1} R =
\left(\begin{matrix} \hat{t}^*_{14} & 0 \cr
c^{-1}_{14} \hat{t}^*_{24} & c^{-1}_{24} \hat{t}^*_{25} \cr
0 & c^{-1}_{25} \hat{t}^*_{35} \cr
\end{matrix}\right) \; .
\label{24}
%     (24)
\end{eqnarray}
In this case the results of $J^{}_\nu$ and $\varepsilon^{}_{1,2}$
can be directly obtained from Eqs.~(\ref{22}) and (\ref{23})
with the replacements $\hat{s}^*_{14} \to \hat{t}^*_{14}$,
$\hat{s}^*_{24} \to c^{-1}_{14} \hat{t}^*_{24}$,
$\hat{s}^*_{25} \to c^{-1}_{24} \hat{t}^*_{25}$ and
$\hat{s}^*_{35} \to c^{-1}_{25} \hat{t}^*_{35}$. It turns out that
these three CP-violating quantities are all proportional to
$\sin 2\beta$ and thus proportional to one another. But such
a simple proportionality between $J^{}_\nu$ and $\varepsilon^{}_{1,2}$
is highly subject to the two-zero texture of $R$ under consideration,
and hence it should not be expected in general.

\subsection{\bf Scenario B: flavor mixing democracy}

Now we assume that all the six active-sterile flavor mixing angles
$\theta^{}_{ij}$ (for $i = 1, 2, 3$ and $j = 4, 5$) are equal to $\theta^{}_0$
in a simplified minimal seesaw scenario, but their associated CP phases
$\delta^{}_{ij}$ should be different from one another to assure that
$\alpha$, $\beta$ and $\gamma$ are not all vanishing. In this case only
the coefficients $C^{}_0$, $C^\prime_{\alpha\beta}$, $C^\prime_{\beta\gamma}$
and $C^\prime_{\gamma\alpha}$ in Eq.~(\ref{15}) can be nonzero, leading us to
\begin{eqnarray}
{\cal J}^{}_\nu \hspace{-0.2cm} & = & \hspace{-0.2cm}
48 \frac{M^2_1 M^2_2 \left(M^2_2 - M^2_1\right)}
{\Delta^{}_{21} \Delta^{}_{31} \Delta^{}_{32}}
s^{12}_0 \sin\frac{\alpha - \beta}{2} \sin\frac{\beta - \gamma}{2}
\sin\frac{\gamma - \alpha}{2}
\nonumber \\
\hspace{-0.2cm} & & \hspace{-0.2cm}
\times \left[\sin^2\frac{\alpha - \beta}{2} + \sin^2\frac{\beta - \gamma}{2}
+ \sin^2\frac{\gamma - \alpha}{2}\right] \; ,
\label{25}
%     (25)
\end{eqnarray}
where $s^{}_0 \equiv \sin\theta^{}_0$ for $\theta^{}_{ij} = \theta^{}_0$.
In comparison, the flavor-independent CP-violating asymmetries
$\varepsilon^{}_1$ and $\varepsilon^{}_2$ for $N^{}_1$ and $N^{}_2$ decays
turn out to be
\begin{eqnarray}
\varepsilon^{}_{1} \hspace{-0.2cm} & = & \hspace{-0.2cm}
+ \frac{M^2_2 \xi(x^{}_{21}) s^2_0}{12\pi \langle \phi^0\rangle^2}
\left(\sin\alpha + \sin\beta + \sin\gamma\right)
\left(\cos\alpha + \cos\beta + \cos\gamma\right) \; ,
\nonumber \\
\varepsilon^{}_{2} \hspace{-0.2cm} & = & \hspace{-0.2cm}
- \frac{M^2_1 \xi(x^{}_{12}) s^2_0}{12\pi \langle \phi^0\rangle^2}
\left(\sin\alpha + \sin\beta + \sin\gamma\right)
\left(\cos\alpha + \cos\beta + \cos\gamma\right) \; . \hspace{0.4cm}
\label{26}
%     (26)
\end{eqnarray}
It is obvious that the dependence of $\varepsilon^{}_1$ and $\varepsilon^{}_2$
on $\alpha$, $\beta$ and $\gamma$ is quite different from that of
${\cal J}^{}_\nu$. The same is true for the dependence of the flavor-dependent
CP-violating asymmetries $\varepsilon^{}_{i \alpha}$ (for $i = 1, 2$ and
$\alpha = e, \mu, \tau$) on $\alpha$, $\beta$ and $\gamma$, as compared with
that of ${\cal J}^{}_\nu$. In other words, it is impossible to achieve a
simple proportional relation like
$\varepsilon^{}_{i \alpha} \propto {\cal J}^{}_\nu$ in the flavor mixing
democracy case. A special example is $\alpha = \beta = \gamma \equiv
\phi^{}_0$, which simply gives rise to $\varepsilon^{}_1 \propto
\varepsilon^{}_2 \propto \sin 2\phi^{}_0$ and ${\cal J}^{}_\nu = 0$.

Scenarios A and B tell us that a careful analysis of the CP-violating
quantities $\varepsilon^{}_{i \alpha}$ and ${\cal J}^{}_\nu$ is absolutely
necessary even in the minimal seesaw framework, because all the special
correlations between them proposed in the literature are strongly
model-dependent (see, e.g., Ref.~\cite{Xing:2020ald}).

A general and explicit seesaw-bridged correlation between CP violation in
light neutrino oscillations and that in heavy Majorana neutrino decays
is certainly more complicated in the canonical case with three
right-handed neutrino fields, but it deserves a comprehensive study in the
precision measurement era of neutrino physics. The relevant results
will be presented in another paper.

\vspace{0.5cm}

{\it The preliminary result of this work was first reported in my talk
given at the International Conference on the Physics of the Two
Infinities, which was held at University of Kyoto from 27 to 30 March 2023.
My recent research has been supported by the National Natural Science
Foundation of China under grant No. 12075254 and grant No. 11835013.}

%\newpage

\end{document}